\documentstyle[eqsecnum,multicol,aps,prl,epsf]{revtex}

\date{\today}
\draft
\tighten
\begin{document}
\def\sqr#1#2{{\vcenter{\hrule height.3pt
      \hbox{\vrule width.3pt height#2pt  \kern#1pt
         \vrule width.3pt}  \hrule height.3pt}}}
\def\square{\mathchoice{\sqr67\,}{\sqr67\,}\sqr{3}{3.5}\sqr{3}{3.5}}
\def\today{\ifcase\month\or
  January\or February\or March\or April\or May\or June\or July\or
  August\or September\or October\or November\or December\fi
  \space\number\day, \number\year}

\def\Bbb{\bf}

\newcommand {\be}{\begin{equation}}
\newcommand {\ee}{\end{equation}}
\newcommand {\bea}{\begin{array}}
\newcommand {\cl}{\centerline}
\newcommand {\eea}{\end{array}}
\renewcommand {\thefootnote}{\fnsymbol{footnote}}

\def\nc{noncommutative }
\def\com{commutative }
\def\ncy{noncommutativity }
\def\repr{representation }
\def\Ham{Hamiltonian }
\def\reps{representations }
\def \simlt{\stackrel{<}{{}_\sim}}
\def \simgt{\stackrel{>}{{}_\sim}}


\title{Hydrogen Atom Spectrum and the Lamb Shift in Noncommutative QED}

\author{M. Chaichian, M.M. Sheikh-Jabbari$^{\dagger}$ and A. Tureanu}

\address {High Energy Physics Division, Department of Physics, University of Helsinki\\ and \\
Helsinki Institute of Physics, FIN-00014 Helsinki, Finland \\
E-mail: Masud.Chaichian@helsinki.fi, atureanu@pcu.helsinki.fi \\
$^{\dagger}$ The Abdus Salam International Center for Theoretical Physics,
Strada Costiera 11,Trieste, Italy\\
E-mail: jabbari@ictp.trieste.it}

\maketitle


\begin{abstract}
We have calculated the energy levels of the hydrogen atom and as
well the Lamb shift within the noncommutative quantum
electrodynamics theory. The results show deviations from the
usual QED both on the classical and on the quantum levels. On
both levels, the deviations depend on the parameter of
space/space noncommutativity.

\end{abstract}

\pacs{PACS: 11.15.-q, 11.30.Er, 11.25.Sq.
{\qquad} {\qquad} $\ \ \ \ $
IC/2000/157, HIP-2000-55/TH, hep-th/0010175}
\vspace*{0.1cm}

\begin{multicols} {2}


\section{Introduction}
\setcounter{equation}{0}

Recently, remotivated by string theory arguments, noncommutative spaces (Moyal plane) have been studied
extensively. The \nc space can be realized by the coordinate operators satisfying
\be\label{NC}
[{\hat x}_{\mu},{\hat x}_{\nu}]=i\theta_{\mu\nu}\ ,
\ee
where ${\hat x}$ are the coordinate operators and $\theta_{\mu\nu}$ is the \ncy parameter and is of dimension of
(length)$^2\ $; for a review on the string theory side, see \cite{SW}.
The action for field theories on \nc spaces, NCFT's, is then obtained using the Weyl-Moyal correspondence
\cite{{AlW},{Micu},{Ihab}}, according to which, in order to find the \nc action, the usual product of fields should
be replaced by the star-product:

\be\label{star}
(f*g)(x)=exp({i\over 2}\theta_{\mu\nu}\partial_{x_{\mu}}\partial_{y_{\nu}})
f(x)g(y)|_{x=y}\ ,
\ee
where $f$ and $g$ are two arbitrary infinitely differentiable functions on $R^{3+1}$.
Performing explicit loop calculations, for $\theta^{0i}=0$ cases (\nc space), it has been shown that \nc
$\phi^4$ theory up to two loops \cite{{Micu},{Aref}} and NCQED up to one loop \cite{{Haya},{Ihab},{Martin}}, are
renormalizable. For \nc space-time ($\theta^{0i}\neq 0$) it has been shown that the theory is not unitary
and hence, as a field theory, it is not appealing \cite{Mehen}.

Apart from the field theory interests which are more academic, we are more interested in  some
possible phenomenological consequences of \ncy in space.
Some of those results, all from the field theory point of view, have been addressed in
\cite{{Ihab},{Roiban},{chair}}. However, perhaps a better starting point is to study quantum
mechanics (QM) on such \nc spaces.
To develop the NCQM formulation we need to introduce a Hamiltonian which governs the time evolution of the
system. We should also specify the phase space and, of course, the Hilbert space on which these operators act.
As for the phase space, inferred from the string theory \cite{{AAS},{Chu}}, we choose
\be\label{NCXP}
\bea{cc}
\left [\hat{x}_i,\hat{x}_j \right ]=i\theta_{ij}\ , \\
\left [\hat{x}_i,\hat{p}_j\right ]=i{\hbar}\delta_{ij}\ , \\
\left [\hat{p}_i,\hat{p}_j\right ]=0\ .
\eea\ee

The Hilbert space can consistently be taken to be exactly the same as the Hilbert space of the
corresponding commutative system.
This assumption for the Hilbert space is directly induced from the non-relativistic limit of the related
NCFT, and one can really argue that it satisfies all the needed properties of a physical Hilbert space.
The only non-trivial part of such a formulation is to give the Hamiltonian. Once we have done it, the dynamical
equation for the state $|\psi\rangle$ is the usual Schrödinger equation, i.e. $H|\psi\rangle=i{\hbar}
{\partial\over \partial t} |\psi\rangle$.

In this letter we focus on the hydrogen atom and, using the
non-relativistic limit of NCQED results, we propose the
Hamiltonian describing the NC H-atom. Given the \Ham and assuming
that the \ncy parameter ($\theta_{ij}$) is small, we study the
spectrum of H-atom. We show that because of noncommutativity,
even at field theory  tree level, we have some corrections to the
Lamb shift ($2P_{1/2}\rightarrow 2S_{1/2}$ transition). Since the \ncy in
space violates rotational symmetry, our Lamb shift corrections
have a preferred direction and hence we call them "polarized
Lamb shift". We also consider further corrections to Lamb shift
originating from the loop contributions in NCQED. In this way we will find 
some upper bound for $\theta$.
In addition, we study the Stark and Zeeman effects for the \nc H-atom.

\section{Formulation of the Noncommutative  Hamiltonian}

To start with, we propose the following \Ham for the \nc H-atom. Of course, we shall verify our proposal by a
NCQED calculation:
\be\label{Ham}
H={{\hat p}.{\hat p}\over 2m}+ V({\hat x})\ ,
\ee
where the Coulomb potential in terms of the noncommutative coordinates
$\hat x$ is:
\be
V(r)=-\frac{Ze^{2}}{\sqrt{\hat{x}\hat{x}}}\ ,
\ee
with ${\hat p}$ and ${\hat x}$ satisfying (\ref{NCXP}).

Now, we note that there is a new coordinate system,

\be\label{com}
x_i=\hat{x}_i+\frac{1}{2\hbar}\theta_{ij}\hat{p}_j\ ,\;\;\;\;\;
p_i=\hat{p}_i\ ,
\ee
where the new variables satisfy the usual canonical commutation
relations:

\begin{eqnarray}
\left [x_i,x_j \right ]&=&0\ ,\cr
\left [x_i,p_j\right ]&=&i{\hbar}\delta_{ij}\ ,\cr
\left [p_i,p_j\right ]&=&0\ .
\end{eqnarray}

So, if in the Hamiltonian we change the variables ${\hat x}_i,\ {\hat p}_i$ to
$x_i,\ p_i$, the Coulomb potential becomes:

\begin{eqnarray}\label{NCppp}
V(r)&=&-\frac{Ze^{2}}{\sqrt{(x_i-\theta_{ij}p_j/2\hbar)(x_i-\theta_{ik}p_k/2\hbar)}}\cr
&=&-\frac{Ze^{2}}{r}-Ze^{2}\frac{x_i\theta_{ij}p_j}{2\hbar
r^{3}}+ O(\theta^2) \cr &=&-\frac{Ze^{2}}{r}-Ze^{2}\frac{L\cdot\theta}{4\hbar
r^{3}}+ O(\theta^2)\ ,
\end{eqnarray}
where $\theta_{ij}=\frac{1}{2}\epsilon_{ijk}\theta_k$, $L=r\times p$.

As $(r\times p)\cdot \theta = - r\cdot (\theta\times p)$, it
follows that the Coulomb potential can be also written as

\be
V(r)=-\frac{Ze^{2}}{r}-\frac{e}{4\hbar}(\theta \times p)\cdot(-\frac{Zer}{r^3})+ O(\theta^2)\ .
\ee
The other higher order terms, besides being  higher powers in $\theta$ which in its own turn is very small,
are also higher powers in momenta.

Our proposal for \Ham can be  justified  from field theory
calculations. The electron-photon vertex function at tree level
in NCQED is \cite{Ihab}:
\be \label{Gamma}
\Gamma_\mu = e^{\frac{i}{2\hbar^{2}}p\times p\prime}\gamma_\mu=
e^{-\frac{i}{2\hbar^{2}}p\cdot \tilde q}\gamma_\mu\ ,  
\ee
where $p$ and $p'$ are the in-coming and out-going electron
momenta, respectively, $q_{\mu}$ is the photon momentum: $p'-p=q$
and
$$
p\times p'=p_i\theta^{ij}p'_j\ ,\;\;\;\;\;\ \tilde q^i=\theta^{ji}q_j\ .
$$

Expanding  the exponential in powers of $\theta$ and keeping only the
first two terms, it appears that the second term will give rise
to an {\it electric dipole moment} \cite{Dip}, which couples to an external
electric field $E$ as $-\langle P\rangle \cdot E$, where

\be \langle P_i\rangle = -\frac{1}{2\hbar}e\tilde p_i =
\frac{1}{2\hbar}e\theta_{ij}p_j\ . 
\ee 
This electric dipole moment, as we will see, changes the usual Lamb shift.
Actually one can go further and prove that the potential (\ref{NCppp}), for all orders in $\theta$, is 
expected from the NCQED starting from (\ref{Gamma}). This can be done noting that
$f(x_i+\epsilon_i)=e^{\epsilon_i{\partial\over \partial x_i}}f(x)$.

Our proposal for the NC H-atom \Ham can be generalized to other systems, i.e. taking the
usual \Ham but now being a function of \nc coordinates (like (\ref{Ham})). However, our discussion
based on NCQED is only applicable when we deal with the "electro-magnetic" interaction. In other words, at field
theory tree level and in the non-relativistic limit, the \ncy of space is probed through the electric dipole
moment of particles, whether fermions or bosons.

In our formulation for NCQM, one can still use the usual definition for the probability density, $|\psi|^2$. However,
one should be aware that there is no coordinate basis in this case. In our approach, since the \ncy
parameter, if it is non-zero, should be very small compared to the length scales of the system, one can always
treat the \nc effects as some perturbations of the \com counter-part and hence, up to first order in $\theta$,
we can use the usual wave functions and probabilities.

\section{"Classical" spectrum for hydrogen atom in NC theory}

Using the usual perturbation theory, the leading corrections to
the energy levels due to noncommutativity, i.e. first order
perturbation and in field theory tree level, are:
\be\label{leading} 
\Delta E_{NC}^{H-atom} = -\langle{nl'jj'_z}|\frac
{Ze^2}{4\hbar} \frac{L\cdot\theta}{r^3}|{nljj_z}\rangle\ . 
\ee 
We note that the above expression is very similar to that of the
spin-orbit coupling, where ${\theta\over  \lambda_e^2}$ is now replacing the
spin, ${\vec{S}\over \hbar}$, with $\lambda_e$ being the electron Compton wave length.

If we put $\theta_3=\theta$ and the rest of the $\theta$-components to zero (which can be done by a rotation or a
redefinition of coordinates), then $L\cdot\theta =L_z\theta$ and,
taking into account the fact that 
$$
\langle{ljj_z}|L_z|{l'jj'_z}\rangle = j_z \hbar (1\mp{1\over 2l+1})\delta_{ll'}\delta_{j_zj'_z}, 
\;\; j=l\pm {1\over 2},\, 
$$
the energy level shift given by (\ref{leading})
becomes \cite{Gas} :

\be\label{NCLamb}
\Delta E_{NC}^{H-atom}=
-{m_ec^2\over 4} (Z\alpha)^4{\theta\over \lambda_e^2}
j_z (1\mp{1\over 2l+1})
f_{n,l}\ \delta_{ll'}\delta_{j_zj'_z}
\ee
for $j=l\pm {1\over 2}$ and $f_{n,l}=\frac{1}{n^{3}l(l+\frac{1}{2})(l+1)}$.

It is worth noting that in order to find $\langle {1\over r^3}\rangle$, one should integrate over
the wave functions from $r=0$; on the other hand, the approximation we are working in (dropping
the terms higher order in $\theta$) is not valid for $r\simlt \sqrt\theta$. However,
since the wave function for $l\neq 0$ is zero at $r=0$, the result (\ref{NCLamb}) still holds at
this level.

The case of our interest, the $2P_{1/2}\rightarrow 2S_{1/2}$ transition
(Lamb shift), differs from the usual \com case in which the shift depends
only on the $l$ quantum number and all the corrections are due to
the field theory loop effects. The Lamb shift for the \nc H-atom, besides the usual loop effects,
depends on the $j_z$ quantum number (only for the $2P_{1/2}$ level, as the levels with $l=0$ are not affected) and is 
there, even in the 
field
theory tree level. Hence we call it {\it polarized
Lamb shift}. More precisely, there is a {\it new} transition
channel which is opened because of noncommutativity:
$2P_{1/2}^{-1/2}\rightarrow 2P_{1/2}^{1/2}$, with the notation $nl_{j}^{j_z}$ for the energy levels. The
usual Lamb shift, $2P_{1/2}\rightarrow 2S_{1/2}$, is now split into two parts,
$2P_{1/2}^{1/2}\rightarrow 2S_{1/2}$ and $2P_{1/2}^{-1/2}\rightarrow 2S_{1/2}$, which means
that the \ncy effects increase the widths and split the Lamb shift line
by a factor proportional to $\theta$.

\section{One-loop corrections}

In the usual \com theory, the Lamb shift is believed to come from {\it loop corrections} to QED.
In the usual case, both vertex corrections, in particular, the $g-2$ factor in the spin-orbit coupling, and the
corrections to
photon  propagator \cite{Zuber} are responsible for the Lamb shift. 
\begin{center}
{\bf A. Noncommutative one loop vertex corrections}
\end{center}
\vskip .2cm
According to NCQED one loop results, the electric and magnetic dipole moments of the electron, as a Dirac particle,
are
\cite{Ihab}:
\be\bea{lll}\label{dipole}
\langle \vec{\mu} \rangle =-{e\over 2m c}(g \vec{S}+ {\alpha\gamma_E\over 3\pi} {\hbar}{\vec{\theta}\over \lambda_e^2})
\;\;\ ,\ g=2+{\alpha\over\pi}\ , \cr 
\langle \vec{P} \rangle =-{e\over 4{\hbar}}(\vec{\theta}\times\vec{p})(1+
\frac{3\alpha\gamma_E}{\pi})\ .
\eea\ee
Hence, the {\it \nc} one loop correction to the potential (\ref{NCppp}), originating from {\it vertex} corrections up
to the first order in $\theta$, is:
\be\label{1Lver}
V^{1Loop}_{\rm{NC\ vertex}}=-{Ze^2\over 4\pi}\gamma_E\alpha(3-{2\over 3}){\vec{L}.\vec{\theta}\over \hbar r^3}\ .
\ee
\begin{center}
{\bf B. Noncommutative one loop photon propagator corrections}
\end{center}
\vskip .2cm
The photon propagator at one loop in the NCQED, for small 
$q, \tilde q$ is given by \cite{Haya}:
$$
\Pi^{\mu \nu}(q)= \frac{e^2}{16 \pi^2}
\bigg[\frac{10}{3}(g^{\mu\nu}q^2-q^{\mu}q^{\nu})\big(\ln{(q^{2}\tilde
q^{2})}+{2\over 25} {q^2\over m^2}\big)
$$ 
\be\label{prop}
+32\frac{\tilde q^{\mu}\tilde q^{\nu}}{\tilde q^4}-\frac{4}{3}\frac{q^2}{\tilde q^2}\tilde q^{\mu}\tilde
q^{\nu}\bigg]\ , 
\ee
where the term proportional to ${2\over 25} {q^2\over m^2}$ is the fermionic loop contribution which, because of
the cancellation in phase factors coming from \ncy, is the same as the usual QED result.   
From (\ref{prop}), by taking only the part of the propagator
corresponding to time-like photons and reintroducing ${\hbar}, c$ factors, in the units where the Coulomb potential is
$-{Ze^2\over r}$, we obtain
$$
V^{1Loop}_{\rm{prop.}}(r)= -Ze^{2}\alpha\frac{10}{3\hbar}\int d^3q{\frac{1}{\vec{q}^{\
2}}e^{-i \vec{q}\cdot\vec{r}/\hbar}}
$$
\be
\;\;\;\;\;\;\;\;\;\;\ \times\bigg(\ln{(\frac{\vec{q}^{\ 2}\tilde
q^{2}}{\hbar^{4}})}+{2\over 25}{\vec{q}^{\ 2}\over m^2}\bigg)\ .
\ee
The second term in the integral yields the usual $\delta^3(r)$ type correction to the Coulomb potential. 
To work out the integral in the first term which is $\theta$ dependent, let us assume that only $\theta_{12}=\theta$ is
non-zero. If we denote the integral by $I(r,\theta)$, then 
$\frac{dI}{d\theta}=\frac{1}{2\pi r\theta}$ and thus $I(\theta, r)= \frac{1}{2\pi
r}\ln{({\theta\Lambda^2})}$, where $\Lambda$ is a cut-off. This can be
understood noting that, because of IR/UV mixing \cite{{Ihab},{Haya}}, the Fourier
transformation and also (\ref{prop}), are valid for  
${1\over\Lambda\theta}\simlt q\simlt \Lambda$. 
Putting all these results together, we have
\begin{eqnarray}\label{1Lprop} 
V^{1Loop}_{\rm{prop.}}(r) =-{Ze^{2}\over {2 \pi r}}\frac{10\alpha}{3}\ln{({\theta
\Lambda^2})}-Ze^2 {4\alpha\over 15} \lambda_e^2 \delta^3(r)\ . 
\end{eqnarray}
The first term being
proportional to ${1\over r}$, can be understood as the normalization of charge at one loop 
level \cite{Haya}; however, to find the physical value of $\alpha$ (NCQED coupling), one should
study the Thomson limit of Compton scattering \cite{Abarb} for the \nc case \cite{prog}. 
Summing up all the one loop contributions to Lamb shift due to noncommutativity,
(\ref{1Lver}), (\ref{1Lprop}), we get
$$ \Delta E_{NC}^{1Loop}=-{1\over 2\pi}m_e c^2 (Z\alpha)^2\bigg[{5\alpha\over {3}}\ln{({\theta\Lambda^2})}{1\over
n^2}-{(Z\alpha)^2\over 2} {\theta\over \lambda_e^2} $$ 
\be
\times\gamma_E\alpha(3-{2\over 3})\frac{j_z(1\mp{1\over 2l+1})}{ n^3 l(l+{1\over2})(l+1)}\bigg]\ . 
\ee 
One can use the data on the Lamb shift to
impose some bounds on the value of the \ncy parameter, $\theta$. Of course, to do it, we only need to consider the
classical (tree level) results, (\ref{NCLamb}). Comparing these results, the contribution of (\ref{NCLamb}) should be
of the order of $10^{-6}-10^{-7}$ smaller than the usual one loop result and hence, $$ {\theta\over \lambda_e^2}\simlt
10^{-7}\alpha \ \ \ \ {\rm or}\ \ \ \theta\simlt (10^4\ GeV)^{-2}\ . $$ This bound is indeed not a strong one, and one
would need some more precise experiments or data. Among other processes, the $e^{+}e^-$ scattering data can provide a
better bound on $\theta$ \cite{prog}.

\section{Noncommutative Stark and Zeeman effects}

\begin{center}
{\bf Stark effect}
\end{center}

The potential energy of the atomic electron in an external
electric field oriented along the $z$-axis is given, at tree level,
by
\be V_{Stark}=eEz+\frac{e}{4\hbar}(\theta\times p)\cdot E \ee
(neglecting the motion of the proton).

The change in the hydrogen atom energy levels due to
noncommutativity (the second term in (5.1)) is:
\be 
\Delta E_{Stark}^{NC}=\langle{nl'jj'_z}|\frac{e}{4\hbar}(\theta\times p)\cdot
E|{nljj_z}\rangle\ , \ee

Taking into account the fact that $p_i=\frac{m}{i\hbar}[x_i,H_0]$,
where $H_0$ is the unperturbed Hamiltonian, so that
$H_0|nljj_z\rangle=E_n|nljj_z\rangle$, the correction to the energy
levels becomes: 
\be
\Delta E_{Stark}^{NC}
\propto (\theta\times E)_i
\langle{nl'jj'_z}|[x_i,H_0]|{nljj_z}\rangle=0 ,
\ee
meaning that, at tree level, the contribution to the Stark effect
due to noncommutativity is zero.
We also note that, adding the one loop corrections to electric dipole moment (\ref{dipole}), the above
result will not be changed.

\begin{center}
{\bf Zeeman effect}
\end{center}

The new parts which are added to the usual potential energy of the atom in a magnetic field, due to noncommutativity,
are:

\be 
V_{{\rm NC\ Zeeman}} =
\frac{e}{2m_{e}c}\frac{\alpha\gamma_{E}m_{e}^{2}}{3\pi\hbar}(1-f
\frac{m_p}{m_e})\vec{\theta}\cdot\vec{B}\ .
\ee 
where $f$ is a form factor of the order of unity, as the proton is
not point-like. As a result, the noncommutative contribution to
the Zeeman effect in the first order of perturbation theory is: 
\begin{eqnarray}
\Delta E_{Zeeman}^{NC} 
=\frac{1}{6\pi c\hbar}e\alpha\gamma_{E}m_{e}(1-f\frac{m_p}{m_e})
\vec{\theta}\cdot\vec{B}\ .
\end{eqnarray} 

\section {Conclusion}

We have presented the results on the classical Coulomb potential within the formulated
noncommutative quantum mechanics for the hydrogen atom and have obtained the corrections to
the Lamb shift using the NCQED. If there exists any
noncommutativity of space-time in nature, as it seems to emerge from different theories and 
arguments, its implications should appear in physical systems such as the one treated in this
letter. A detailed analysis of the results obtained here, together with the treatment of
other fundamental and precisely measured physical processes, 
will be given in a further communication \cite {prog}.

We are grateful to A. Demichev, D. Demir, P. Pre\v{s}najder and especially to C. Montonen
for many useful discussions and comments. This work was partially supported by the Academy of
Finland, under the Project No. 163394. The work of M.M. Sh.-J. was partly supported by the EC
contract no. ERBFMRX-CT 96-0090.


\end{multicols}
\end{document}